# Cosmological number density n(z) in depth z from p(V/V$_m$) distribution

Dilip G. Banhatti

School of Physics, Madurai Kamaraj University, Madurai 625021

**Abstract.** Using distribution p(V/V$_m$) of V/V$_m$ rather than just mean <V/V$_m$> in V/V$_m$-test leads directly to cosmological number density n(z). Calculation of n(z) from p(V/V$_m$) is illustrated using best sample (of 76 quasars) available in 1981, when method was developed. This is only illustrative, sample being too small for any meaningful results.

## Luminosity-distance and volume

For cosmological populations of objects, distance is measured by (monochromatic) luminosity-distance $\ell_\nu(z)$ (at frequency ν), function of redshift z of object. Similarly, volume of sphere passing through object and centered around observer is $(4.\pi/3).v(z)$. Both $\ell_\nu(z)$ and v(z) are specific known functions of z for given cosmological model.

## Calculation of limiting redshift z$_m$

For source of (monochromatic radio) luminosity L$_\nu$, flux density S$_\nu$, (radio) spectral index α
(α ≡ - dlog S$_\nu$ / dlog ν), and redshift z, L$_\nu$ = 4.π.$\ell_\nu^2$(α, z).S$_\nu$. For survey limit S$_0$, value of limiting redshift z$_m$ is given by $\ell_\nu^2$(α, z) / $\ell_\nu^2$(α, z$_m$) = S$_0$ / S$_\nu$ ≡ s, 0 ≤ s ≤ 1, for source of redshift z and spectral index α. For simplest case,
[$\ell_\nu$(α, z) / $\ell_\nu$(α, z$_m$)]$^2$ = s has single finite solution z$_m$ for given α, z and S$_\nu$, S$_0$. Different values z$_m$ correspond to different L$_\nu$(α).

## Relating n(z) to p(V/V$_m$)

Let N(z$_m$).dz$_m$ represent number of sources of limiting redshifts between z$_m$ and z$_m$ + dz$_m$ in sample covering solid angle ω of sky. Then 4.π.N(z$_m$) / ω is total number of sources of limit z$_m$ per unit z$_m$-interval. Since volume available to source of limit z$_m$ is
V(z$_m$) = (4.π / 3).(c / H$_0$)$^3$.v(z$_m$), (where speed of light c and Hubble constant H$_0$ together determine linear scale of universe,) number of sources (per unit z$_m$-interval) per unit volume is
{3.N(z$_m$) / ω}.(H$_0$ / c)$^3$.(1 / v$_m$), where v$_m$ ≡ v(z$_m$). Let n$_m$(z$_m$, z) be number of sources / unit volume / unit z$_m$-interval at redshift z. Then, n(z) ≡ $\int_z^\infty$ dz$_m$. n$_m$(z$_m$, z), and
n$_m$(z$_m$, z) = {3.N(z$_m$) / ω}.(H$_0$ / c)$^3$.(1 / v(z$_m$)).p$_m$(v(z) / v(z$_m$)) for 0 ≤ z ≤ z$_m$, where p$_m$(x) is distribution of x ≡ V/V$_m$ for given z$_m$. For z > z$_m$, n$_m$(z$_m$, z) = 0, since sources with limiting redshift z$_m$ cannot have z > z$_m$. To get n(z) for all z$_m$-values, integrate over z$_m$:
n(z) = {3 / ω}.(H$_0$ / c)$^3$. $\int_z^\infty$ dz$_m$.( N(z$_m$) / v(z$_m$)).p$_m$(v(z) / v(z$_m$)).

## Scheme of Calculation

Any real sample has maximum z$_{max}$ for z$_m$. So, n(z$_{max}$) = 0. In fact, lifetimes of individual sources will come into consideration, as well as structure-formation epoch at some high redshift
(say, z > 10). Thus, n(z) calculation will give useful results only upto redshift much less than z$_{max}$.
Formally writing z$_{max}$ instead of ∞ for upper limit,
n(z) = {3 / ω}.(H$_0$ / c)$^3$. $\int_z^{z\_max}$ dz$_m$.( N(z$_m$) / v(z$_m$)).p$_m$(v(z) / v(z$_m$)) for 0 ≤ z ≤ z$_{max}$.
To apply to real samples, this must be converted to sum. Divide z$_m$-range 0 to z$_{max}$ into k equal intervals, each = z$_{max}$ / k = Δz. Mid-points are
z$_j$ = (j − ½).Δz = {(j − ½) / k}.z$_{max}$. Calculate n(z) at these points: n(z$_j$). Converting integral to sum,
(ω / 3).(c / H$_0$)$^3$.n(z$_j$) = $\sum_{i=j}^{k}$ {N$_i$ / v(z$_i$)}.p$_i$(x$_{ij}$), where x$_{ij}$ = v(z$_j$) / v(z$_i$).                    **(1)**

# Illustrative Calculation in 1981

Wills & Lynds (1978) have defined carefully sample of 76 optically identified quasars. We use this sample only to illustrate derivation of $n(z)$ from $p(x) \equiv p(V/V_m)$. We use Einstein-de Sitter cosmology or $q_0 = \sigma_0 = \tfrac{1}{2}$, $k = \lambda_0 = 0$ or $(\tfrac{1}{2}, \tfrac{1}{2}, 0, 0)$ world model in von Hoerner's (1974) notation, for which

$(H_0/c)^2 . \ell_v^2(\alpha, z) = 4.(1+z)^\alpha / \{\sqrt{(1+z)} - 1\}^2$ and $(H_0/c)^3 . v(z) = 8.\{1 - 1/\sqrt{(1+z)}\}^3$.

For each quasar, $z_m$ is calculated by iteration with initial guess $z$ for $z_m$. Values of $z$, $z_m$ are then used to calculate $v(z)$, $v(z_m)$ and hence $x = V/V_m$. All 76 $V/V_m$-values are used to plot histogram. Good approximation for $p(x)$ is $p(x) = 2.x$, which is normalized over [0,1]. The limiting redshifts $z_m$ range from 0 to 3.2. Dividing into four equal intervals, bins centered at 0.4, 1.2, 2.0 and 2.8 contain 19, 31, 16 and 10 quasars. Although each of these 4 subsets is quite small, we calculate and plot histograms $p_i(x)$, $i = 1, 2, 3, 4$ for each subset for x-intervals of width 0.2 from 0 to 1, thus with 5 intervals centered at $x = 0.1, 0.3, 0.5, 0.7$ and $0.9$. Each normalized $p_i(x)$ is also well approximated by $p_i(x) = 2.x$ except $p_4(0.2994)$. So we do calculations using this approximation in addition to using actual values. Finally we calculate $(\omega / 3).(c/H_0)^3.n(z_j)$ using **(1)**. (See tables.)

### Table for $p_i(x)$ and $p(x)$

| X | No. | $p_1(x)$ | No. | $p_2(x)$ | No. | $p_3(x)$ | No. | $P_4(x)$ | No. | $p(x)$ |
|---|---|---|---|---|---|---|---|---|---|---|
| 0.1 | 0 | 0 | 1 | 0.161 | 0 | 0 | 0 | 0 | 1 | 0.066 |
| 0.3 | 2 | 0.526 | 2 | 0.323 | 3 | 0.9375 | 1 | 0.5 | 8 | 0.526 |
| 0.5 | 3 | 0.789 | 6 | 0.968 | 2 | 0.625 | 1 | 0.5 | 12 | 0.789 |
| 0.7 | 8 | 2.105 | 8 | 1.290 | 7 | 2.1875 | 5 | 2.5 | 28 | 1.842 |
| 0.9 | 6 | 1.580 | 14 | 2.258 | 4 | 1.25 | 3 | 1.5 | 27 | 1.776 |
| Totals | 19 | | 31 | | 16 | | 10 | | 76 | |

### Table of $n(z)$ calculation

| j | $z_j$ | $N_j$ | → $v(z_j)$ | i = 1 | i = 2 | i = 3 | i = 4 | → $n(z_j)$ |
|---|---|---|---|---|---|---|---|---|
| 1 | 0.4 | 19 | 2.97E-2 | 1 | 0.1074 | 0.0492 | 0.0321 | 1307. |
| 2 | 1.2 | 31 | 0.27666 | | 1 | 0.4580 | 0.2994 | 255. |
| 3 | 2.0 | 16 | 0.60399 | | | 1 | 0.6536 | 67. |
| 4 | 2.8 | 10 | 0.92407 | | | | 1 | 22. |

Notes for second table: (a) 5$^{th}$ to 8$^{th}$ columns list $x_{ij}$-values,
(b) → $v(z_j) \equiv (H_0/c)^3.v(z_j) = 8.\{1 - 1/\sqrt{(1+z_j)}\}^3$, and
(c) → $n(z_j) \equiv (\omega/3).(c/H_0)^3.n(z_j)$.
Use of approximations $p_i(x) = 2.x$ in evaluating sums **(1)** for each row j = 1, 2, 3, 4 gives virtually same results.

## Conclusion
Due to too small sample, results are only indicative. Main aim is illustrating method fully.

## Acknowledgments

Work reported evolved out of discussions with Vasant K Kulkarni in 1981. Computer Centre of IISc, Bangaluru was used for calculations. First draft was written in 2004-2005 in Muenster, Germany. Radha D Banhatti provided, as always, unstinting material, moral & spiritual support. Uni-Muenster is acknowledged for use of facilities & UGC, New Delhi, India for financial support.

-------------------------------------------------------------------------------------------------------------------